# Direct experimental evidence of tunable charge transfer at the LaNiO$_3$/CaMnO$_3$ ferromagnetic interface

J. R. Paudel[1], M. Terilli[2], T.-C. Wu[2], J. D. Grassi[1], A. M. Derrico[1], R. K. Sah[1], M. Kareev[2], C. Klewe[3], P. Shafer[3], A. Gloskovskii[4], C. Schlueter[4], V. N. Strocov[5], J. Chakhalian[2], and A. X. Gray[1,*]

[1] *Physics Department, Temple University, Philadelphia, Pennsylvania 19122, USA*
[2] *Department of Physics and Astronomy, Rutgers University, Piscataway, New Jersey 08854, USA*
[3] *Advanced Light Source, Lawrence Berkeley National Laboratory, Berkeley, California 94720, USA*
[4] *Deutsches Elektronen-Synchrotron, DESY, 22607 Hamburg, Germany*
[5] *Swiss Light Source, Paul Scherrer Institute, 5232 Villigen, Switzerland*
*email: axgray@temple.edu*

## Abstract

Interfacial charge transfer in oxide heterostructures gives rise to a rich variety of electronic and magnetic phenomena. Designing heterostructures where one of the thin-film components exhibits a metal-insulator transition opens a promising avenue for controlling such phenomena both statically and dynamically. In this letter, we utilize a combination of depth-resolved soft X-ray standing-wave and hard X-ray photoelectron spectroscopies in conjunction with polarization-dependent X-ray absorption spectroscopy to investigate the effects of the metal-insulator transition in LaNiO$_3$ on the electronic and magnetic states at the LaNiO$_3$/CaMnO$_3$ interface. We report on a direct observation of the reduced effective valence state of the interfacial Mn cations in the metallic superlattice with an above-critical LaNiO$_3$ thickness (6 u.c.) due to the leakage of itinerant Ni 3$d$ e$_g$ electrons into the interfacial CaMnO$_3$ layer. Conversely, in an insulating superlattice with a below-critical LaNiO$_3$ thickness of 2 u.c., a homogeneous effective valence state of Mn is observed throughout the CaMnO$_3$ layers due to the blockage of charge transfer across the interface. The ability to switch and tune interfacial charge transfer enables precise control of the emergent ferromagnetic state at the LaNiO$_3$/CaMnO$_3$ interface and, thus, has far-reaching consequences on the future strategies for the design of next-generation spintronic devices.

Application-driven atomic-level design of complex-oxide heterostructures that exhibit functional electronic and magnetic phenomena has become a diverse and vibrant subfield of condensed matter physics and materials science [1-3]. Of particular interest are the materials systems wherein rich physics and intricate interplay between various degrees of freedom at the interface give rise to functional properties not observed in the constituent materials [4-8]. In such heterostructures, charge transfer across the interface often plays a key role in establishing new electronic [9-11], magnetic [12,13], and orbital [14-16] states with properties that can be tailored via dimensionality [17,18], epitaxial strain [19,20], interface termination [21,22], doping [23,24], and engineered defects [25,26]. Thus, the flexibility and diversity of the perovskite oxide structures coupled with state-of-the-art thin-film synthesis turn the fundamental and robust phenomenon of charge transfer into a powerful tuning knob for creating the desired ground state and controlling its functionality.

Epitaxial superlattices consisting of antiferromagnetic $CaMnO_3$ and paramagnetic $LaNiO_3$ exhibit emergent ferromagnetism [13] that can be tailored by varying the thickness of individual layers [27]. Such thickness dependence has been attributed to the electronic-structural changes in the $LaNiO_3$ layer, which undergoes a metal-insulator transition in the ultrathin (few-unit-cell) limit [28]. According to our current understanding based on several experimental studies [13,27,29,30] and a theoretical study of the similar $CaMnO_3/CaRuO_3$ system [31], in metallic superlattices with an above-critical $LaNiO_3$ thickness, interfacial charge transfer mediated by the itinerant Ni $3d$ $e_g$ electrons is expected to create an increased concentration of $Mn^{3+}$ cations at the interface. Such charge reconstruction creates an electronic environment favorable for the emergence of the $Mn^{4+}$- $Mn^{3+}$ double exchange interaction, which stabilizes long-range canted ferromagnetic order in an approximately one-unit-cell-thick interfacial layer of $CaMnO_3$. Conversely, in superlattices with a below-critical $LaNiO_3$ thickness (<4 u.c.), partial or complete blockage of charge transfer from the now-insulating $LaNiO_3$ results in a significant (approximately three-fold) suppression of the observed magnetic moment [27]. The residual magnetic moment of approximately 0.3 µB per interfacial Mn atom has been attributed to the presence of the $Ni^{2+}$-$Mn^{4+}$ superexchange interaction made possible by the oxygen vacancies in the $LaNiO_3$ layers [27,30] that are driven to the $LaNiO_3$/$CaMnO_3$ interfaces by the polar mismatch. Such defects, as well as possible chemical intermixing, play a minor role and could be controlled via growth conditions and epitaxial strain [26,32]. Thus, the phenomenon of interfacial charge transfer is currently considered to be the main driving force responsible for the emergence of the long-range ferromagnetic order at the $LaNiO_3$/$CaMnO_3$ interface. Furthermore, the possibility of switching and tuning this magnetic phenomenon in the quantum confined structure via either static or dynamic control of the metallicity of the $LaNiO_3$ layers makes the $LaNiO_3$/$CaMnO_3$ system a prime candidate for high-density spintronic devices wherein energy-efficient magnetic switching could be accomplished with electric fields or other external stimuli.

Detection and characterization of interfacial charge transfer phenomena like the ones described above present a unique practical challenge due to the lack of direct yet non-destructive techniques capable of probing minute changes in the valence state at a buried interface with element specificity and Ångstrom-level spatial resolution. The $LaNiO_3$/$CaMnO_3$ heterostructure exemplifies a class of materials systems where such stringent measurement requirements are necessary due to the extremely localized nature of the phenomenon of interest.

In this letter, we report on a direct observation of the tunable character of interfacial charge transfer in $LaNiO_3/CaMnO_3$ superlattices using depth-resolved soft X-ray standing-wave photoelectron spectroscopy (SW-XPS) [33]. We examined two otherwise identical superlattices containing either insulating or metallic $LaNiO_3$ layers and extracted element-specific and valence-state-sensitive spectroscopic data from the interfacial regions. Standing-wave (SW) excitations in both first-order and second-order Bragg reflection geometries were utilized to enhance the depth resolution of the technique. Our results revealed a depth-dependent modification of the effective valence state on the interfacial Mn cations in the metallic [6 u.c. $LaNiO_3$ / 4 u.c. $CaMnO_3$]×10 superlattice that is consistent with the enhanced charge-transfer picture. Conversely, a homogeneous effective valence state throughout the $CaMnO_3$ layers was observed for the thinner [2 u.c. $LaNiO_3$ / 4 u.c. $CaMnO_3$]×10 superlattice, suggesting suppression of charge transfer across the interface due to the insulating nature of the below-critical-thickness $LaNiO_3$ films. Complementary bulk-sensitive hard X-ray photoelectron spectroscopy (HAXPES) measurements of the valence-band electronic structure revealed suppression of the Ni $3d$ $e_g$ density of states near the Fermi level, consistent with the thickness-dependent metal-insulator transition in the $LaNiO_3$ layers, which was confirmed via conventional electronic transport measurements. Concomitant suppression of the interfacial magnetic moment on the interfacial Mn sites was observed via polarization-dependent X-ray absorption spectroscopy (XAS) with X-ray magnetic circular dichroism (XMCD) at the Mn $L_{2,3}$ edges.

For our experiments, two high-quality epitaxial superlattices consisting of ten repetitions of $LaNiO_3/CaMnO_3$ were synthesized using pulsed laser interval deposition [34] on a single-crystalline $LaAlO_3$(001) substrate. The thickness of the $CaMnO_3$ layers in the superlattices was kept the same at 4 u.c., while the thickness of the $LaNiO_3$ layers was fixed at N = 2 u.c. for the first (insulating) superlattice and N = 6 u.c. for the second (metallic) superlattice. Thicknesses and layer-by-layer deposition were monitored *in situ* using reflection high-energy electron diffraction (RHEED). After the growth, the resultant high quality, crystallinity, and correct layering of the superlattices were confirmed *ex situ* using X-ray diffraction (XRD). Correct chemical composition was confirmed via bulk-sensitive HAXPES measurements carried out using a laboratory-based spectrometer equipped with a 5.41 keV monochromated X-ray source and a Scienta Omicron EW4000 high-energy hemispherical analyzer. Individual thicknesses of the layers and the quality of the interfaces were confirmed using synchrotron-based SW-XPS measurements that are described in detail later in this letter (see Fig. 2). The results of the XRD, RHEED, and HAXPES characterization are presented in Figures S1 and S2 of the Supplementary Information.

To probe the thickness-dependent variation in the valence-band electronic structure of $LaNiO_3$ and its effect on the resistivity of the superlattices, we utilized a combination of bulk-sensitive valence-band HAXPES spectroscopy and electronic transport measurements. The bulk-sensitive valence-band HAXPES measurements were carried out at the P22 beamline [35] of the PETRA III synchrotron (DESY) using a photon energy of 6.0 keV. At this excitation energy, the values of the inelastic mean-free path (IMFP) of the valence-band photoelectrons in $CaMnO_3$ and $LaNiO_3$ are estimated to be 87 Å and 71 Å, respectively, with the maximum probing depth being approximately three times these values [36]. The total experimental energy resolution (380 meV at the analyzer pass energy of 50 eV) and the position of the zero binding energy were determined by measuring the Fermi edge of a standard Au sample. The measurements were carried out at the sample temperature of approximately 60 K.

Figure 1a shows the experimental valence-band spectra of the N = 6 u.c. superlattice (red line) and the N = 2 u.c. superlattice (blue line). The corresponding temperature-dependent sheet

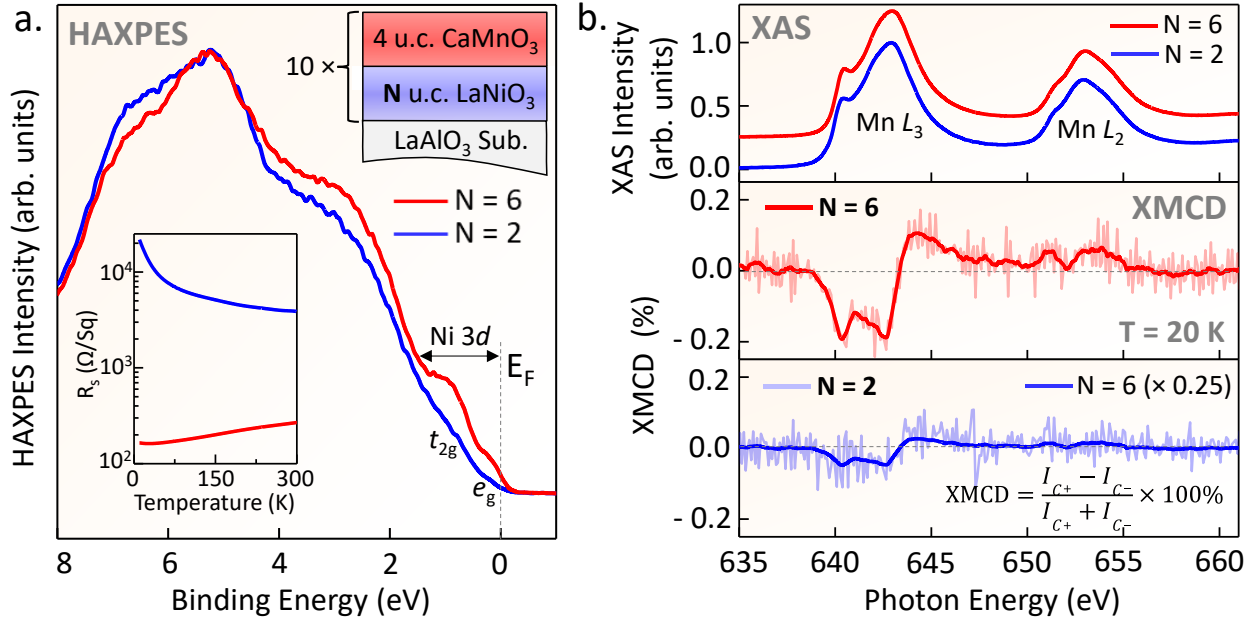

**Figure 1. Metal-insulator transition and suppression of the interfacial ferromagnetism. a.** Angle-integrated bulk-sensitive HAXPES valence-band spectra of the N = 6 u.c. (red line) and N = 2 u.c. (blue line) superlattices recorded with a photon energy of 6 keV at T = 60 K. Significant depletion of the near-$E_F$ Ni 3$d$ $e_g$ and $t_{2g}$ density of states results in a metal-insulator transition in the N = 2 u.c. sample, as probed with sheet resistivity measurements shown in the inset. **b.** Bulk-sensitive Mn $L_{2,3}$ edge XAS spectra measured in an LY detection mode at T = 20 K and probing the entire depth of the superlattice (top panel) reveal no significant differences in the depth-averaged valence states of Mn between the two samples. The XMCD spectrum of the N = 6 u.c. superlattice (middle panel) shows a significant magnetic signal of up to -0.20% at the Mn $L_3$ edge. The light-red spectrum represents raw data while the solid red curve has been smoothed using the Savitzky-Golay method. Conversely, the N = 2 u.c. spectrum (light-blue spectrum in the bottom panel) exhibits a nearly negligible XMCD signal of approximately -0.05% in the same photon energy range. The solid blue spectrum represents the N = 6 data scaled by a factor of four (×0.25), for comparison.

resistance curves measured using the standard van der Pauw method are shown in the inset. The near-Fermi-level region of the spectra exhibits two prominent features which, based on prior studies, corresponds to the strongly-hybridized Ni 3$d$ $e_g$ and $t_{2g}$ states at 0.3 eV and 1.0 eV, respectively [28,37]. It is important to note that the spectral range near the Fermi level does not contain any CaMnO$_3$-derived electronic states due to its insulating nature with a bandgap

"window" on the order of 1.35 - 3.05 eV as determined via theory and experiment, respectively [38,39]. Consistent with prior studies of the thickness-dependent metal-insulator transition in LaNiO$_3$ [28,37], the superlattice containing below-critical-thickness LaNiO$_3$ layers (N = 2 u.c.) exhibits a significant suppression of the near-Fermi-level electronic states resulting directly in the approximately two orders of magnitude enhancement in sheet resistivity. In the context of interfacial ferromagnetism, such bulk-sensitive HAXPES measurements demonstrate directly the depletion of the itinerant Ni 3$d$ e$_g$ states that, in metallic superlattices, facilitate charge transfer from Ni to the interfacial Mn sites, thus creating an electronic environment that stabilizes the ferromagnetic state mediated by the double exchange interaction.

The effect of suppressing the Ni-to-Mn charge transfer channel in the N = 2 u.c. superlattice is immediately evident in the element-specific (Mn) XMCD measurements of these two otherwise identical samples. The XAS measurements at the Mn $L_{2,3}$ edges (top panel of Fig. 1b) do not exhibit significant differences between the two superlattices, due to the depth-averaging nature of the technique. On the other hand, the XMCD spectra shown in the lower panels are only sensitive to the interfacial ferromagnetic state and, therefore, isolate the magnetic signal from the interface (without facilitating quantitative depth resolution). Consistent with prior studies [13,27], the thicker N = 6 u.c. superlattice exhibits a significant XMCD signal (approximately -0.20% at the Mn $L_3$ edge), which is directly proportional to the magnetic moment on Mn in the interfacial region of the CaMnO$_3$ layers. Conversely, the below-critical-thickness N = 2 u.c. superlattice exhibits almost no discernable XMCD signal throughout the spectrum, except a slight excursion from zero (approximately -0.05%) at the photon energies corresponding to the Mn $L_3$ absorption threshold (640-643 eV). Such a four-fold suppression of the Mn magnetic moment is qualitatively consistent with a prior study by Flint *et al.* [27], where magnetic moments of up to 1 µB/Mn were observed for the above-critical-thickness superlattices and only 0.2 µB/Mn for the ultrathin (2 u.c. LaNiO$_3$) samples. All of our XAS and XMCD measurements were carried out at the high-resolution (100 meV) Magnetic Spectroscopy beamline 4.0.2 at the Advanced Light Source [40] in the bulk-sensitive luminescence yield (LY) detection mode at T = 20 K.

Our measurements carried out using bulk-sensitive spectroscopies such as HAXPES and XAS/XMCD (LY), in concert with prior studies suggest a direct connection between the depletion of the Ni 3$d$ e$_g$ states near the Fermi level in the ultrathin (N = 2 u.c.) LaNiO$_3$ films and the suppression of interfacial magnetic moment on Mn sites in the adjacent CaMnO$_3$ films. In order to bridge this connection and link it to interfacial charge transfer from Ni to Mn, we have measured the depth-resolved evolution of the Mn valence state at the LaNiO$_3$/CaMnO$_3$ interface in both samples using soft X-ray standing-wave photoemission spectroscopy (SW-XPS).

Our experiments were performed at the soft-X-ray ARPES endstation [41] of the high-resolution (95 meV) ADRESS beamline at the Swiss Light Source [42] equipped with a SPECS PHOIBOS-150 hemispherical electrostatic analyzer and a six-axis cryogenic manipulator at T = 20 K. The measurements were carried out using a p-polarized X-ray beam in the grazing incidence geometry with a footprint on the sample measuring 75 µm by 32 µm along the vertical and horizontal axes, respectively. The photoelectron detection was done in near-normal-emission geometry with the analyzer acceptance angle set to ±8⁰, parallel to the analyzed slit [41]. The results are presented in Figures 2 and 3 below.

In the SW-XPS technique, shown schematically in Figure 2a, Ångstrom-level depth resolution is facilitated by generating an X-ray SW interference field within a periodic superlattice sample [33,43]. The maximum-contrast modulations in the X-ray *E*-field intensity are achieved at the first- and second-order Bragg conditions that are typically found by varying the X-ray grazing incidence angle at a fixed photon energy. Once the X-ray SW field is established within the sample, it can be translated vertically (perpendicular to the sample's surface) by approximately half of the superlattice period by scanning (rocking) the grazing X-ray incidence angle across the Bragg condition. In a recent soft X-ray study, the depth resolution of one cubic perovskite unit cell (approximately 3.8 Å) has been demonstrated using the same experimental setup [44].

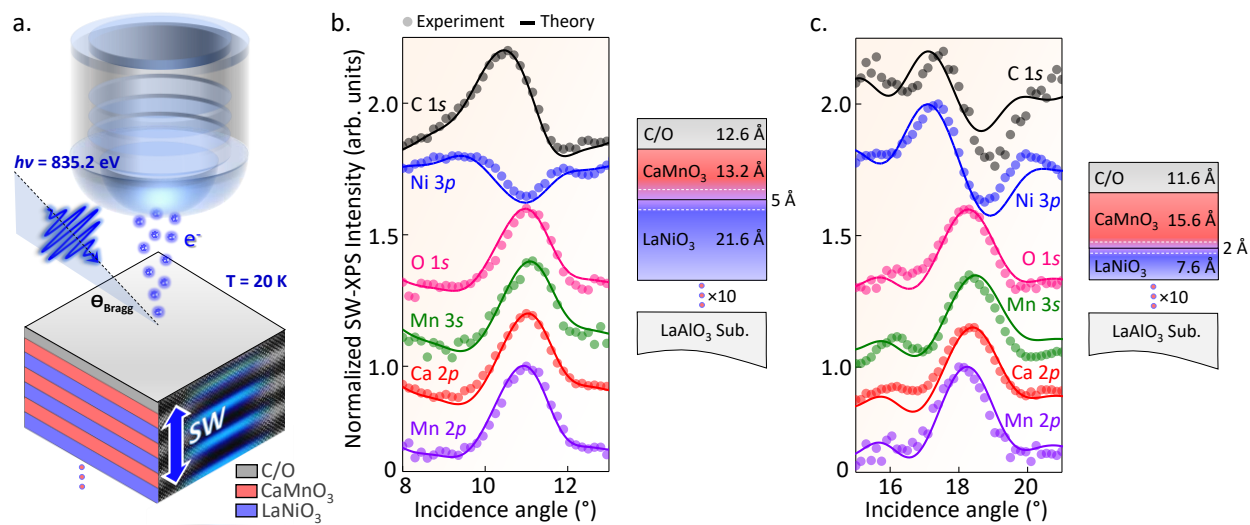

**Figure 2. SW-XPS elemental depth profiling and X-ray optical modeling of the superlattices. a.** Schematic diagram of the SW-XPS experiment and the investigated superlattice structures consisting of 10 LaNiO$_3$/CaMnO$_3$ bilayers grown epitaxially on a LaAlO$_3$(001) substrate, with each bilayer consisting of 4 u.c. of CaMnO$_3$ and N u.c. (N = 6 and 2) of LaNiO$_3$. **b.** Best fits between the experimental (circular markers) and calculated (solid lines) SW rocking curves for all accessible representative core levels in the thicker (N = 6 u.c.) superlattice. The resultant depth profile yielding the values of the individual layer thicknesses and interface roughness (interdiffusion) is shown on the right. **c.** Similar results of the X-ray optical fitting of the experimental SW rocking curves for the thinner (N = 2 u.c.) superlattice and the resultant depth profile.

As the first step in such an experiment, incidence-angle-dependent 'rocking curves' of the core-level intensities for several constituent elements in the superlattice are measured in the first-order Bragg condition for each sample. Typically, for unambiguous X-ray optical fitting, it is necessary to record and analyze such rocking curves for at least one element from each layer, as well as the adventitious carbon from the surface atmospheric contaminant layer [45].

Figures 2b and 2c show experimental rocking-curve spectra for the integrated intensities of the Mn 2$p$, Mn 3$s$, Ca 2$p$, O 1$s$, Ni 3$p$, and C 1$s$ core-level peaks measured on the N = 6 u.c. and N = 2 u.c. superlattices, respectively (circular markers). The measurements were carried out with a resonant photon energy of 835.2 eV (La 3$d_{5/2}$ absorption threshold) to maximize the X-ray optical contrast between LaNiO$_3$ and CaMnO$_3$ [46]. It is immediately obvious that the Mn/Ca, Ni, and C signals originate from different layers (vertical locations) within the sample due to the contrasting lineshapes (phases) of their respective rocking-curve spectra. The O 1s rocking curve is dominated by the signal from the upper CaMnO$_3$ layer due to the limited probing depth and, therefore, resembles the Ca and Mn spectra, consistent with prior studies [30,44]. The Bragg features for the thinner (N = 2 u.c.) superlattice appear at higher grazing incidence angles (16º-20º), as expected from the basic diffraction formalism.

The experimental rocking curves were simultaneously and self-consistently fitted using X-ray optical theoretical code for automatic structure optimization [47]. The code is based on the formalism developed by Yang *et al.* [48] that accounts for the differential photoelectric cross sections of relevant orbitals and the photoelectron effective attenuation lengths (EAL), calculated for each layer. Thicknesses of the CaMnO$_3$ and LaNiO$_3$ layers and the interface roughness (interdiffusion length) were constrained to be uniform throughout the superlattice and allowed to vary in the model. Fixed X-ray optical constants measured using XAS (imaginary part) and calculated via the Kramers-Kronig transformation (real part) were used for the fitting.

The best fits (solid curves) to the experimental data and the resultant Ångstrom-level depth-resolved chemical profiles of the superlattices are shown in Figures 2b-c. The resultant total thicknesses of the 4 u.c. CaMnO$_3$ and 6 u.c. LaNiO$_3$ layers in the thicker superlattice are consistent to within 0.5 u.c with the lattice constants of 3.89 Å (LaNiO$_3$) and 3.73 Å (CaMnO$_3$), reported previously in the literature [49-51]. For the thinner (N = 2 u.c.) superlattice, the resultant total thicknesses are consistent within <0.2 u.c. The best-fit values of the interface roughness (interdiffusion length) are 5 Å (~1.3 u.c.) for the thicker superlattice and 2 Å (~0.5 u.c.) for the thinner superlattice, which is consistent with typical high-quality layer-by-layer growth [34]. Finally, the thickness of the surface-adsorbed contaminant layer from exposure to the atmosphere (labeled C/O) has typical values of 12-13 Å. Such an element-specific structural analysis adds to the host of other characterization results attesting to the high quality and precise control of our superlattice synthesis.

The X-ray optical models and sample structures were then used to calculate the depth-resolved profiles of the X-ray SW electric field intensities ($E^2$) inside the two superlattices in the first-order Bragg condition. The purpose of such calculations is to determine the two X-ray grazing incidence angles for each of the superlattices, wherein the antinodes (high $E^2$) of the standing wave preferentially highlight (1) the middle "bulk-like" sections of the CaMnO$_3$ layers and (2) the interfacial sections adjacent to LaNiO$_3$. Consequently, by scanning the X-ray grazing incidence angle between these two values, a detailed center-to-interface valence-state profile on Mn can be obtained by measuring the depth-resolved evolution of the Mn 3$s$ core-level multiplet splitting [30].

The results of such calculations for the experimental geometries corresponding to the first-order Bragg condition are shown in Figure S3 in the Supporting Information. It is evident that while the optimal experimental geometries highlighting the bulk-like and interface-like sections of the CaMnO$_3$ layers can be attained for the thinner (N = 2 u.c.) sample (see Fig. S3b), the tripled thickness of the LaNiO$_3$ layers (N = 6 u.c.) and the nearly-doubled period (10 u.c.) of the thicker superlattice precludes such depth-selective measurements with comparable contrast and depth resolution in the first-order Bragg condition (see Fig. S3a).

It has been suggested by Libera *et al.* [52] that carrying out X-ray SW experiments in the second-order Bragg geometry could result in an enhancement of the depth resolution due to the doubling of the SW frequency within the sample (at the expense of some diminution of the SW contrast). Thus, for the thicker superlattice (N = 6 u.c.) with nearly double the period (10 u.c. compared to 6 u.c.), such SW frequency doubling within the sample results in a similar SW intensity profile and, therefore, facilitates a comparable depth resolution to that seen in the thinner sample (measured in the first-order Bragg condition).

The simulated SW *E*-field intensity profiles for the optimal X-ray grazing incidence geometries highlighting the bulk-like and interface-like sections of the CaMnO$_3$ layers are shown in Figure 3. Figure 3a depicts the second-order SW profiles in the topmost layers of the thicker (N = 6 u.c.) superlattice. It reveals that the optimal X-ray grazing incidence for probing the interface-like and bulk-like sections of the CaMnO$_3$ layers are 23.25⁰ and 25.50⁰, respectively. Conversely, the first-order SW profiles for the thinner (N = 2 u.c.) superlattice shown in Figure 3b reveal the optimal angles of 16.60⁰ (interface) and 19.15⁰ (bulk).

To quantify the valence state of Mn in the bulk CaMnO$_3$ and at the interfaces, we carried out high-resolution measurements of the multiplet-split Mn 3*s* core-level peaks in each of the above-mentioned experimental geometries. In transition-metal oxides, the exchange-coupling interaction between the 3*s* core−hole and 3*d* electrons in the valence bands result in the splitting of the 3*s* core level [53]. The magnitude of this energy splitting is inversely proportional to the valence state of the Mn ion and, thus, can be used to estimate the said valence state [54]. To date, Mn 3*s* core-level energy splitting XPS analysis is considered to be one of the most reliable methods for accurately quantifying changes in the Mn valence state because it does not rely on the relative peak-intensities analysis (*e.g.* XAS and EELS), which can be heavily affected by the background subtraction, limited energy resolution, and the degree of overlap between the peaks corresponding to different valence states (*e.g.* Mn $L_{2,3}$ XAS analysis).

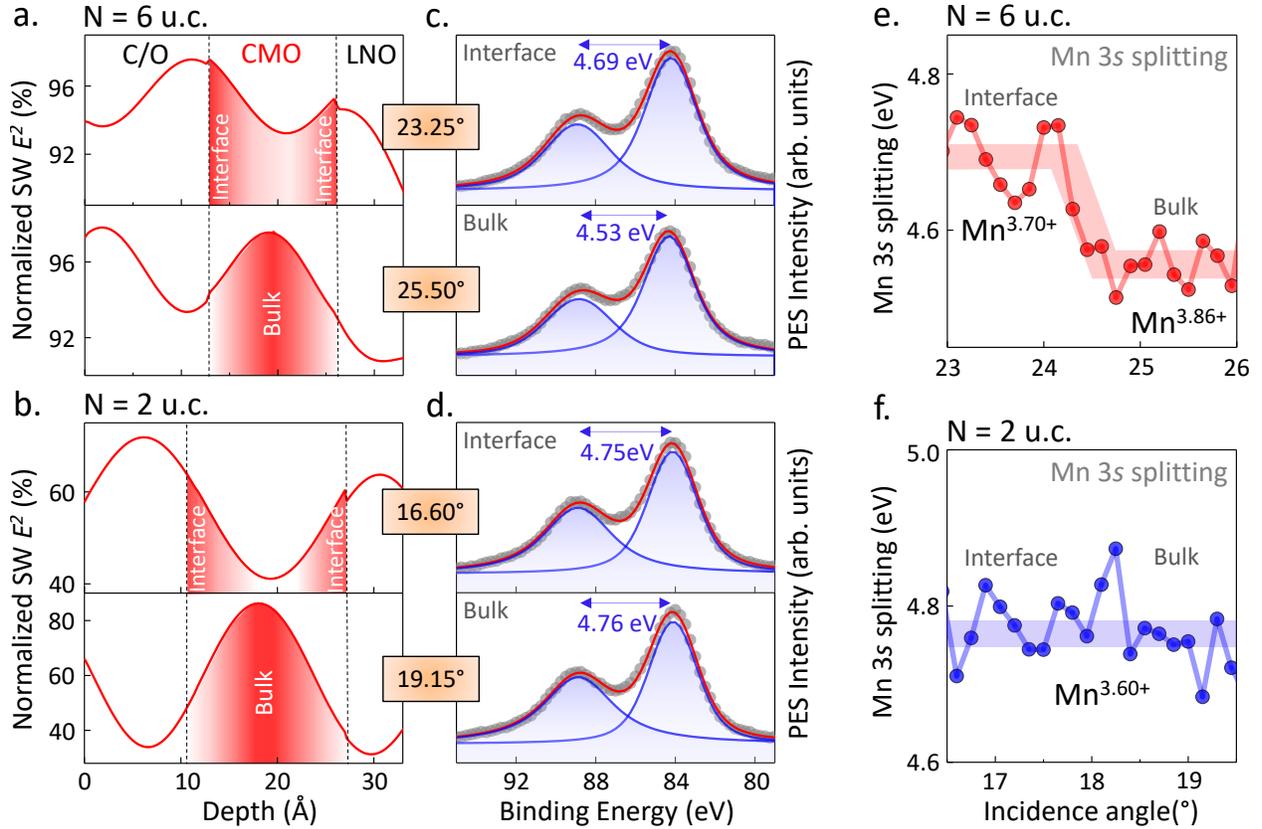

**Figure 3. a.** Simulated intensity of the X-ray standing-wave electric field ($E^2$) inside the N = 6 u.c. superlattice as a function of depth for X-ray grazing incidence angles of 23.25° (top panel) and 25.50° (bottom panel) sensitive to interface-like and bulk-like regions of the CaMnO$_3$ layer, respectively. **b.** Similar simulations for the N = 2 u.c. superlattice with characteristic X-ray incidence angles of 16.60° and 19.15° corresponding to interface-sensitive and bulk-sensitive experimental geometries, respectively. **c.** Mn 3s core-level photoemission spectra for the N = 6 u.c. superlattice recorded in the interface-sensitive (top) and bulk-sensitive (bottom) experimental geometries, respectively. A difference in the multiplet energy splitting of 0.16 eV is observed. **d.** Similar measurements of the Mn 3s core-level spectra for the N = 2 u.c. superlattice reveal no differences in the magnitude of the splitting between the interface and the bulk. **e.** Depth-dependent evolution of the Mn 3s multiplet splitting (in eV) as a function of the X-ray grazing incidence angle showing a change in the effective valence state of Mn in CaMnO$_3$ from +3.86 (in the bulk) to +3.70 (at the interface) in the N = 6 u.c. superlattice. **f.** Similar measurement for the N = 2 u.c. superlattice, showing a homogeneous Mn valence state throughout the thickness of the CaMnO$_3$ layers.

Figures 3c and 3d show Mn 3*s* peaks for the N = 6 u.c. and N = 2 u.c. superlattices, respectively, measured and fitted self-consistently using two Voigt lineshapes after a Shirley background subtraction. The two panels in each figure correspond to the interface-like and bulk-like experimental geometries characterized by the X-ray grazing incidence angle and the SW *E*-field intensity profile shown on the left of the plot.

It is clear that even with the lower SW contrast for the second-order Bragg geometry, a significant increase in the magnitude of the Mn 3*s* energy splitting (~160 meV) for the N = 6 u.c. superlattice is observed at the interface ($\Delta E$ = 4.69 eV) as compared to bulk-like $CaMnO_3$ ($\Delta E$ = 4.53 eV). These values of splitting can be used to estimate the formal valence state of the Mn cation in the bulk and at the interface of the 4 u.c.-thick $CaMnO_3$ film [54], which yields values of +3.86 (near-stoichiometric +4) for the bulk and +3.70 (reduced by 0.16 $e^-$) at the interface.

It is important to note that the observed bulk-to-interface change in the Mn 3*s* multiplet splitting (~160 meV) is approximately 70% larger than our total experimental energy resolution (95 meV). However, since the two multiplet components of the Mn 3*s* peaks are well-separated by nearly 5 eV, our ability to quantify the central positions of the corresponding Voigt peaks is much better than 95 meV and is more closely related to the energy stability of the beamline (<10 meV).

In contrast to this, for the thinner (N = 2 u.c.) superlattice, the magnitudes of the Mn 3*s* core-level multiplet splitting are virtually identical at 4.75 eV (interface) and 4.76 eV (bulk). This result suggests that the valence state of Mn cations in the thinner (N = 2 u.c.) superlattice does not evolve with depth or as a function of proximity to the interface with $LaNiO_3$. Due to a higher SW contrast in the first-order Bragg geometry, our measurement for the thinner superlattice is even more sensitive to depth-dependent changes. It is important to reiterate here that the only difference between the two superlattices investigated in this study is the thickness (and thus metallicity) of the LaNiO$_3$ layers. The thickness of the CaMnO$_3$ layers (4 u.c.) as well as the number of layers in the superlattice (10) are the same for both samples.

To further confirm our findings and to investigate the depth-dependent evolution of the Mn valence state in both samples, we scanned the X-ray grazing incidence angle in 0.2⁰ steps between the two 'extreme' SW conditions, effectively translating the SW antinode from the interface to the central section of the 4 u.c.-thick CaMnO$_3$ film. The resultant plots of the Mn 3*s* core-level multiplet splittings (in eV) as a function of the X-ray grazing incidence angle are shown in Figures 3e and 3f using the same relative vertical (energy) scales.

For the thicker (N = 6 u.c.) superlattice (Fig. 3e), a clear steplike decrease in the magnitude of the splitting is observed over the range of X-ray grazing incidence angles spanning the second-order Bragg condition (24⁰-25⁰). In this narrow angular range, the intensity contrast of the SW within the sample is maximized, and its phase changes (shifts) from interface-sensitive to bulk-sensitive. These findings suggest that the formal valency of the Mn cations changes from approximately +3.86 to +3.70 over a distance of 2 u.c. from bulk to interface, respectively. Thus, the interfacial unit cells of CaMnO$_3$ in the superlattice with thicker (metallic) LaNiO$_3$ layers host an increased concentration of Mn$^{3+}$ cations compared to bulk. The electronic (charge-transfer) nature of this phenomenon is strongly evidenced by the absence of such interfacial valency reduction in the superlattice with the below-critical-thickness (insulating) LaNiO$_3$ layers, as shown in Figure 3f.

It is important to note that some minor but measurable oscillation in the magnitude of the Mn 3*s* splitting is observed at both lower (interface-sensitive) and higher (bulk-sensitive) X-ray grazing incidence angles. Such oscillations are typically observed in SW-XPS measurements of core-level intensities as well as X-ray reflectivity and are termed the Kiessig (or Fresnel) fringes [46]. The intensity modulations result in modulations of the SW contrast (or amplitude). Thus, at certain X-ray grazing incidence angles (*e.g.* 23.7º), we observe some averaging between the interface-like and bulk-like Mn 3*s* energy splitting values.

Previous studies have suggested that the reduced effective valence state of the interfacial Mn cations in metallic superlattices with an above-critical LaNiO$_3$ thickness occurs due to charge transfer of itinerant Ni 3*d* e$_g$ electrons to Mn in the interfacial CaMnO$_3$ layer [13,27,30]. The resultant charge reconstruction at the interface creates an electronic environment favorable for the emergence of the Mn$^{4+}$-Mn$^{3+}$ double exchange interaction, which stabilizes a long-range canted ferromagnetic order. Conversely, in the insulating superlattices with below-critical LaNiO$_3$ thickness, the depletion of the Ni 3*d* e$_g$ states at the Fermi level results in a partial or complete blockage of charge transfer from Ni to Mn and thus prevents the stabilization of the ferromagnetic state. Prior to this study, the amount of charge transfer necessary to stabilize the ferromagnetic order in CaMnO$_3$ had not been measured directly but was calculated to be in the wide range of 0.07-0.20 *e*$^-$ per Mn cation [31].

Thus, our current results establish a direct connection between the depletion of Ni 3*d* e$_g$ states (measured via HAXPES) that results in the metal-insulator transition in LaNiO$_3$ (measured via electronic transport) and the suppression of the charge-transfer-induced ferromagnetic state in CaMnO$_3$ (measured via Mn *L*$_{2,3}$ XMCD). Furthermore, we directly observe the depth-dependent reduction of the Mn valency by 0.16 *e*$^-$ at the interface with metallic LaNiO$_3$ (Fig. 3e) - an amount that is consistent with the prior theoretical prediction of the charge transfer to Mn in a similar materials system (0.07-0.20 *e*$^-$) [31]. In contrast to this, in a similar superlattice with insulating

LaNiO$_3$, no such change in the interfacial Mn valency is observed (Fig. 3f), consistent with the charge-transfer suppression scenario.

It is important to note that neither of the two samples contains perfectly stoichiometric CaMnO$_3$ with an effective Mn valence state of 4+. This is typically the case with coherently-epitaxial CaMnO$_3$ films under tensile strain due to the formation of oxygen vacancies [26]. The SW-XPS measurements of the thinner (N = 2. u.c.) superlattice (see Fig. 3f) yield a lower average valence state of Mn (+3.60) due to several experimental factors, such as a higher surface-sensitivity of the first-order SW measurement as well as a possibly slightly higher tensile strain at the surface of the thinner superlattice (less relaxation) leading to more oxygen vacancies. Our bulk-sensitive HAXPES measurements of the Mn 3$s$ core-level splitting confirm the same depth-averaged Mn valence state (+3.85) in both superlattices (see Figure S4a in the Supporting Information), which is consistent with the similarly bulk-sensitive XAS (LY) measurements shown in Figure 1b. To confirm the surface-like origin of the lower valence state of Mn in the thinner (N = 2 u.c.) sample, we carried out additional surface-sensitive Mn $L_{2,3}$ XAS measurements using the total electron yield (TEY) detection mode, which facilitates an average probing depth of only ~5 nm [55]. The results are shown in Figure S4c of the Supplementary information and reveal a significant increase in the intensities of the Mn$^{3+}$ features at the lower-photon-energy side of the Mn $L_3$ absorption edge (at ~639-641 eV), thus confirming the surface origin of this effect.

It is also important to note that the strain state of CaMnO$_3$ has been shown to have no measurable effect on interfacial ferromagnetism in prior studies [56]. In fact, a significant interfacial ferromagnetic moment on Mn was observed even for the samples grown in a different crystallographic orientation (111 as opposed to 001) [57]. Thus, any minor structural differences between the two measured samples are not likely to have a significant effect on the electronic and magnetic structure at the interface.

In conclusion, we used a combination of bulk-sensitive valence-band HAXPES, magnetic spectroscopy, and electronic transport measurements to probe layer-resolved electronic and magnetic properties of $LaNiO_3$/$CaMnO_3$ superlattices. Our results established a direct connection between the depletion of the Ni $3d$ $e_g$ states leading to the metal-insulator transition in $LaNiO_3$ and the concomitant suppression of the interfacial ferromagnetic state in $CaMnO_3$. We then utilized depth-resolved SW-XPS in both first- and second-order Bragg reflection geometries to link the emergence of interfacial ferromagnetism in $CaMnO_3$ to the direct observation of Ni-Mn charge-transfer-induced valence-state change (by 0.16 e$^-$) of the interfacial Mn cations in the metallic (N = 6 u.c.) superlattice. The tunable (or switchable) character of this phenomenon was demonstrated by tailoring the thickness of the individual $LaNiO_3$ layers in the high-quality layer-by-layer grown $LaNiO_3$/$CaMnO_3$ superlattices. Our results provide a new recipe for designing next-generation spintronic devices using charge-transfer phenomena for efficient tuning and switching of low-dimensional electronic and magnetic states at interfaces.

**Supporting Information**

Supporting information is available after the References section below.

**Notes:**

The authors declare no competing financial interest.

Correspondence and requests for materials should be addressed to A.X.G.

**ACKNOWLEDGEMENTS**

J.R.P., J.D.G, A.M.D., R.K.S., and A.X.G. acknowledge support from the US Department of Energy, Office of Science, Office of Basic Energy Sciences, Materials Sciences and Engineering Division under award number DE-SC0019297. The electrostatic photoelectron analyzer for the lab-based HAXPES measurements at Temple University was acquired through an Army Research Office DURIP grant No. W911NF-18-1-0251. M.T., T.-C.W., M.K., and J.C. acknowledge support from the Department of Energy under Grant No. DE-SC0012375. C.K. and P.S acknowledge support from the US Department of Energy, Office of Science, Office of Basic Energy Sciences, the Microelectronics Co-Design Research Program, under contract no. DE-AC02-05-CH11231 (Codesign of Ultra-Low-Voltage Beyond CMOS Microelectronics). This research used resources of the Advanced Light Source, which is a DOE Office of Science User Facility under Contract No. DE-AC02-05CH11231. We acknowledge DESY (Hamburg, Germany), a member of the Helmholtz Association HGF, for the provision of experimental facilities. Beamtime at DESY was allocated for proposal I-20210142. Funding for the HAXPES instrument at beamline P22 by the Federal Ministry of Education and Research (BMBF) under framework program ErUM is gratefully acknowledged.

J. R. Paudel[1], M. Terilli[2], T.-C. Wu[2], J. D. Grassi[1], A. M. Derrico[1], R. K. Sah[1], M. Kareev[2], C. Klewe[3], P. Shafer[3], A. Gloskovskii[4], C. Schlueter[4], V. N. Strocov[5], J. Chakhalian[2], and A. X. Gray[1,*]

[1] *Physics Department, Temple University, Philadelphia, Pennsylvania 19122, USA*
[2] *Department of Physics and Astronomy, Rutgers University, Piscataway, New Jersey 08854, USA*
[3] *Advanced Light Source, Lawrence Berkeley National Laboratory, Berkeley, California 94720, USA*
[4] *Deutsches Elektronen-Synchrotron, DESY, 22607 Hamburg, Germany*
[5] *Swiss Light Source, Paul Scherrer Institute, 5232 Villigen, Switzerland*
*email: axgray@temple.edu

**Figure S1: X-ray Diffraction (XRD) and RHEED**

Figures S1a and S1b (next page) show the X-ray diffraction θ-2θ spectra for the $N = 2$ u.c. and $N = 6$ u.c. superlattices, respectively. The (002) LaAlO$_3$ substrate peak is observed at $2\theta = 48°$ for both samples, consistent with prior studies [1,2]. The $0^{th}$-order superlattice peak (SL$_0$) for the thinner ($N = 2$ u.c.) sample is obscured by the substrate peak (Fig S1a). For the thicker ($N = 6$ u.c.) superlattice (Fig S1b), the same peak is observed at a slightly lower angle (47.15°), consistent with prior studies [3,4]. The two $1^{st}$-order superlattice peaks for the thinner ($N = 2$ u.c.) sample are observed at the symmetric angular positions of 44° and 52°, approximately ±4° from the $0^{th}$-order peak. Conversely, for the thicker superlattice, the two $1^{st}$-order superlattice peaks are observed closer to the $0^{th}$-order peak (±2.5°), as expected [3]. All of the superlattice peaks exhibit shapes characteristic of high-quality single-crystalline superlattices. Both superlattices show the expected number ($P − 2 = 8$) of SL thickness fringes for $P = 10$ superlattice periods. Observation of the pronounced SL thickness fringes for both superlattices indicates that the superlattice layers are reasonably smooth. Reflection high-energy electron diffraction (RHEED) patterns recorded immediately after deposition are shown in the insets.

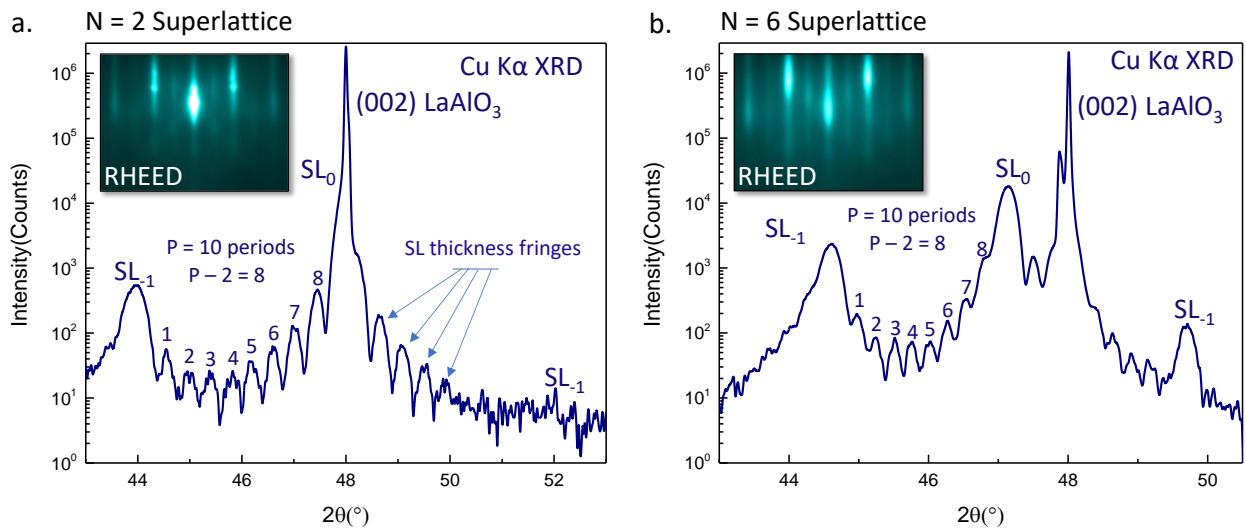

**Figure S1. a.** XRD θ-2θ spectra for the N = 2 u.c. superlattice. The 1$^{st}$-order superlattice peak (SL$_0$) is obscured by the (002) LaAlO$_3$ substrate peak at 2θ = 48°. **b.** XRD θ-2θ spectra for the N = 6 u.c. superlattice. The 1$^{st}$-order superlattice peak (SL$_0$) is observed at 47.15°. Both superlattices show the expected number (P − 2 = 8) of SL thickness fringes for P = 10 superlattice periods. Reflection high-energy electron diffraction (RHEED) patterns recorded immediately after deposition are shown in the insets.

**Figure S2: HAXPES characterization**

The nominal chemical composition of the superlattices was confirmed using bulk-sensitive HAXPES measurements carried out using a laboratory-based spectrometer equipped with a 5.41 keV monochromated X-ray source and a Scienta Omicron EW4000 high-energy hemispherical analyzer. Figure S2 (next page) shows wide-energy range HAXPES survey spectra for the N = 6 u.c. (red line) and N = 2 u.c. (blue line) superlattices. The presence of all expected elements (Ca, Mn, O, La, Ni, and C from the surface-adsorbed contaminant C/O layer) is confirmed by the presence of corresponding core-level peaks. The larger total thickness of the LaNiO$_3$ layers in the red (N = 6 u.c.) spectrum is evidenced by the higher relative intensity of the La 3$d$ and 4$p$ peaks. Measurements were carried out at room temperature.

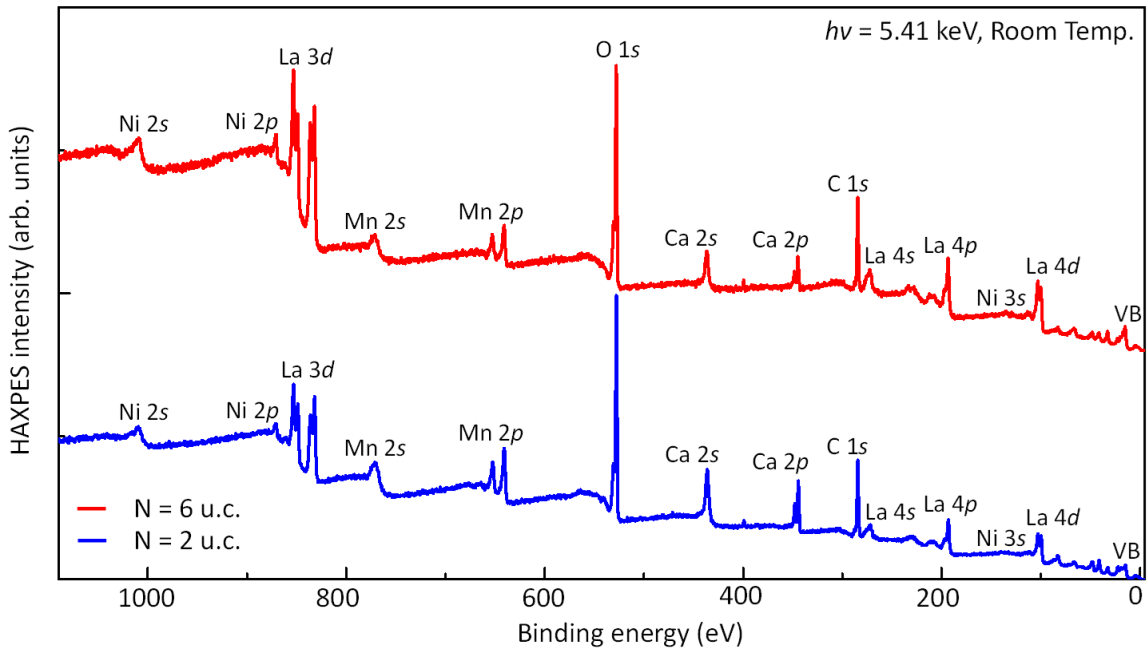

**Figure S2.** HAXPES survey spectra for the N = 6 u.c. (red line) and N = 2 u.c. (blue line) superlattices.

**Figure S3: X-ray SW *E*-field intensity Simulations**

Figures S3a and S3b (next page) show the calculated depth-resolved profiles of the X-ray SW electric field intensities ($E^2$) inside the two superlattices in the first-order Bragg condition. For each superlattice, SW profiles for two X-ray grazing incidence angles are shown, optimized for probing the interfacial regions (top panel) and the central bulklike region (bottom panel) of the $CaMnO_3$ film. It is evident that, while optimal experimental geometries can be attained for the thinner (N = 2 u.c.) sample (Fig. S3b), using the first-order Bragg condition precludes such depth-selective measurements with comparable contrast and depth resolution for the thicker (N = 6 u.c.) sample (Fig. S3a). Specifically, the SW profile at the lowest X-ray grazing incidence angle within the Bragg condition (9.50°) mainly highlights the surface of the sample and suppresses the signal from the $LaNiO_3/CaMnO_3$ interface (top panel). Conversely, for the most bulk-sensitive geometry (12.00°), the nearly-doubled period (10 u.c.) of the superlattice results in a broad $E^2$ peak within the $CaMnO_3$ layer highlighting, once again, the topmost region of the film.

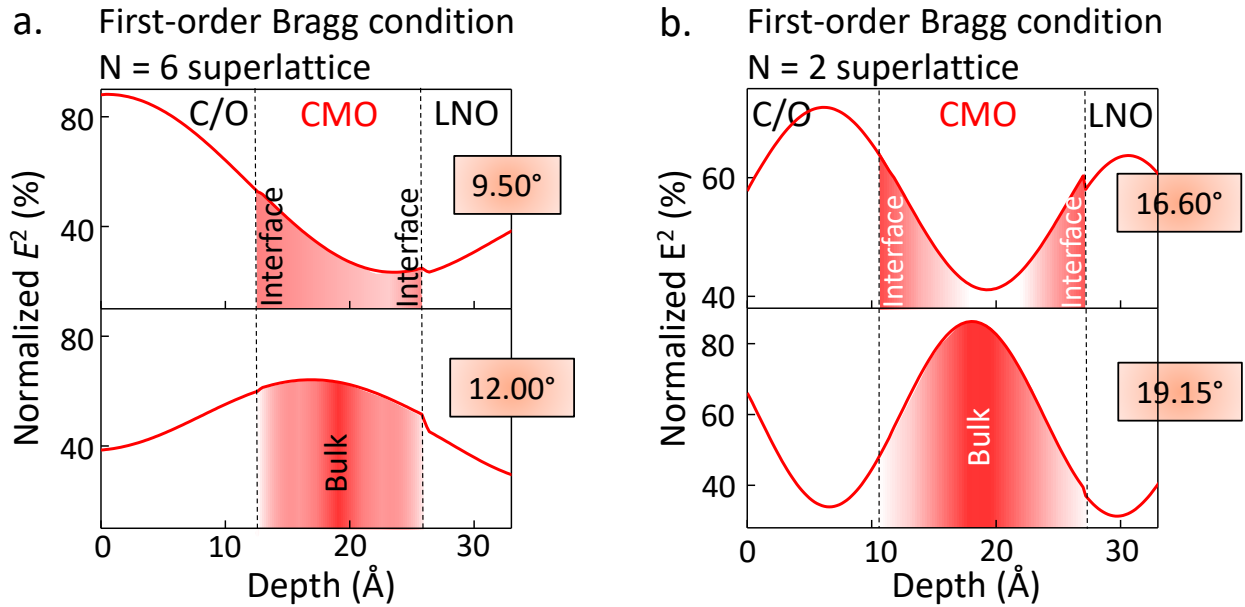

**Figure S3. a.** Simulated intensities of the X-ray standing-wave electric field ($E^2$) inside the N = 6 u.c. superlattice as a function of depth in the first-order Bragg condition. Top panel shows that the interface-sensitive SW profile is not achievable even at the lowest X-ray grazing incidence angle within the Bragg condition (9.50º). The bottom panel shows the SW profile in the optimal bulk-sensitive geometry (12.00º), which is still too 'flat' and primarily highlights the top region of CaMnO$_3$. **b.** Similar simulations for the N = 2 u.c. superlattice, with the characteristic X-ray incidence angles of 16.60° and 19.15° corresponding to the interface-sensitive and bulk-sensitive experimental geometries, respectively (same as Fig. 3b in the main text).

**Figure S4: Surface and bulk-sensitive Mn valence-state analysis**

Figure S4a shows bulk-sensitive HAXPES measurements of the Mn 3$s$ core-levels for the N = 2 u.c. superlattice (blue spectrum) and the N = 6 u.c. superlattice (red spectrum). At the excitation energy of 6.0 keV, the values of the inelastic mean-free path (IMFP) of the Mn 3$s$ photoelectrons in CaMnO$_3$ and LaNiO$_3$ are estimated to be 86 Å and 70 Å, respectively, with the maximum probing depth being approximately three times these values (~23.4 nm) [5]. With this probing depth, we are sensitive to the entire thickness of the superlattice.

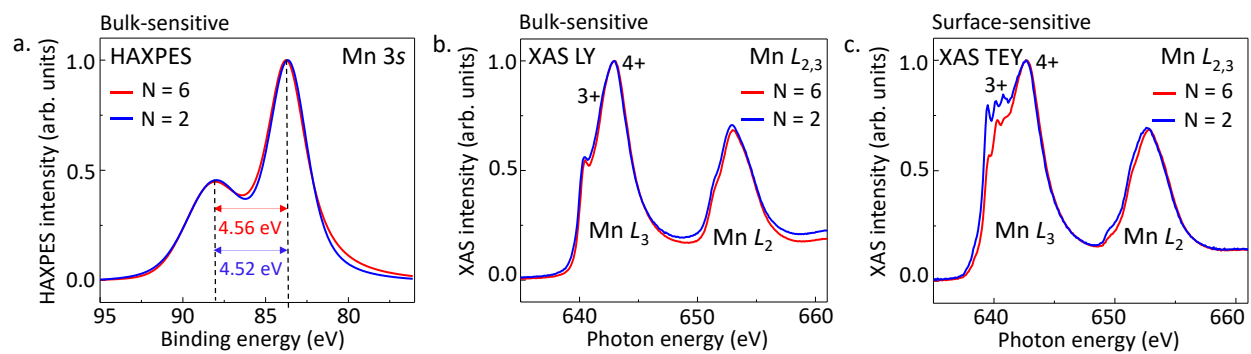

**Figure S4. a.** Bulk-sensitive Mn 3$s$ core-level measurements of the N = 2 u.c. (blue spectrum) and the N = 6 u.c. (red spectrum) superlattices. **b.** Bulk-sensitive XAS LY measurements confirming similar depth-averaged Mn valence states in both superlattices. **c.** Surface-sensitive XAS TEY measurements confirming the surface origin of the $Mn^{3+}$ cations in the N = 2 u.c. superlattice.

Our data reveal that both superlattices exhibit similar depth-averaged Mn 3$s$ multiplet splittings of 4.52 eV (N = 2 u.c.) and 4.56 eV (N = 6 u.c.) that correspond to the effective average Mn valence states of +3.88 and +3.86, respectively [6]. These findings are confirmed using bulk-sensitive Mn $L_{2,3}$ edge XAS spectroscopy with luminescence yield (LY) detection mode, which is sensitive to the entire depth of the superlattice. The bulk-sensitive XAS LY spectra shown in Figure S4b above reveal that the $CaMnO_3$ layers in both superlattices contain Mn that is predominantly in the +4 valence state, with a small fraction of $Mn^{3+}$ characterized by the low-photon-energy shoulder at ~640 eV. XAS spectra recorded in the total electron yield (TEY) detection mode (Fig. S4c) reveal a significant increase in the intensities of the $Mn^{3+}$ features near the surface.